\documentclass[onecolumn,letterpaper,amsmath,amssymb,floatfix,aps,superscriptaddress]{revtex4}

\usepackage{graphicx}
\usepackage{dcolumn}  
\usepackage{bm}  
\usepackage{epsfig}

\begin{document}
\setlength\arraycolsep{2pt}

\title{Sample-to-sample fluctuations of electrostatic forces
generated by quenched charge  disorder}

\author{David S. Dean}
\affiliation{Laboratoire de
Physique Th\'eorique (IRSAMC),Universit\'e de Toulouse, UPS and CNRS,  F-31062 Toulouse, France}

\author{Ali Naji}
\affiliation{Department of Applied Mathematics and Theoretical Physics, Centre for Mathematical Sciences, University of Cambridge, Cambridge 
CB3 0WA, United Kingdom}

\author{Rudolf Podgornik}
\affiliation{Department of Theoretical Physics, J. Stefan
Institute, SI-1000 Ljubljana, Slovenia} 
\affiliation{Institute of
Biophysics, School of Medicine and Department of Physics, Faculty
of Mathematics and Physics, University of Ljubljana, SI-1000
Ljubljana, Slovenia}

\begin{abstract}
It has been recently shown that randomly charged surfaces can exhibit long range 
electrostatic interactions even when they are net neutral. These forces depend
on the specific realization of charge disorder and thus exhibit sample to sample
fluctuations about their mean value. We analyze the fluctuations of these forces
in the parallel slab configuration and also in the sphere-plane geometry via the proximity force approximation. 
The fluctuations of the normal forces, that have a finite mean value, are computed exactly. Surprisingly, we also 
show that lateral forces are present, despite the fact that they have a zero mean, and that their fluctuations 
have the same scaling behavior as the normal force fluctuations.
The measurement of these lateral force fluctuations could help to characterize
the effects of charge disorder in experimental systems, leading to estimates of their
magnitudes that are complementary to those given by normal force measurements.
\end{abstract}
\maketitle

\section{Introduction}

Stability of soft and biological matter in particular is mostly an outcome of the equilibrium between variable range Coulomb and long range van der Waals - Casimir interactions that feature in a plethora of contexts \cite{French}. Though direct measurements of the latter have been announced by various experimental groups the details and the accuracy of experiments are sometimes questioned by the experimentalists themselves \cite{review}. Over the last few years there have been increasing concerns over how to effectively differentiate between the long range Coulomb interactions and the long range van der Waals-Casimir interactions in experiments on interactions between metallic bodies {\sl in vacuo} \cite{kim}. A number of authors have pointed out that disorder effects may significantly affect the forces between surfaces in such experiments; possible sources of disorder include the random surface electrostatic potential \cite{speake} connected  with the so-called patch effect due to the variation of local crystallographic axes of the exposed surface of a clean polycrystalline sample \cite {barrett} as well as effects of disorder in the local dielectric constant \cite{randomdean}. The direct detection of disorder effects in Casimir force experiments, when the force is measured as a function of the intersurface separation of two plates or standardly between a plate and a sphere, is difficult as they must be unravelled from the other forces which are always present \cite{kim}. 

Recently it has been proposed that quenched random charge disorder on surfaces as well as in the bulk can lead to long range interactions even when the surfaces are net-neutral \cite{cd1,cd2}. These long range interactions are induced by a subtle image charge effect and they could play a significant role in experiments to measure the Casimir force as well as in colloidal science in general \cite{Rudi-Ali}, where, for instance, random surface charging can occur during preparation of surfactant coated surfaces \cite{surf}.  Other examples of objects bearing random charge are random polyelectrolytes and polyampholytes \cite{ranpol}. 
While in the latter case the charge distribution could be quenched (i.e., intrinsic to the chain assembly during the polymerization process) or annealed 
(i.e., when  monomers have weak acidic or basic groups that can charge
 regulate depending on the $p$H of the solution), it is not unequivocal to asses the nature of the charge disorder distribution in the case of surfactant coated surfaces.
 
 In this paper we show that net-neutral surfaces experience two types of disorder generated forces that thus show pronounced sample-to-sample fluctuations. The first disorder generated force is normal to the interacting surfaces, whose features we have already investigated in detail elsewhere \cite{cd1}, and shows a non-zero average and fluctuations proportional to the average. The second one, addressed in detail here, is the lateral disorder generated force, acting within the plane parallel to that of the interacting surfaces, whose average is zero but nevertheless exhibits sample-to-sample fluctuations which can be quite large. In principle measurements of lateral force fluctuations could be useful in characterizing and unravelling the effects of quenched charge disorder and thus help the analysis of its role in normal force measurements. 

\begin{figure}[t!]
\begin{center}
\includegraphics[angle=0,width=14cm]{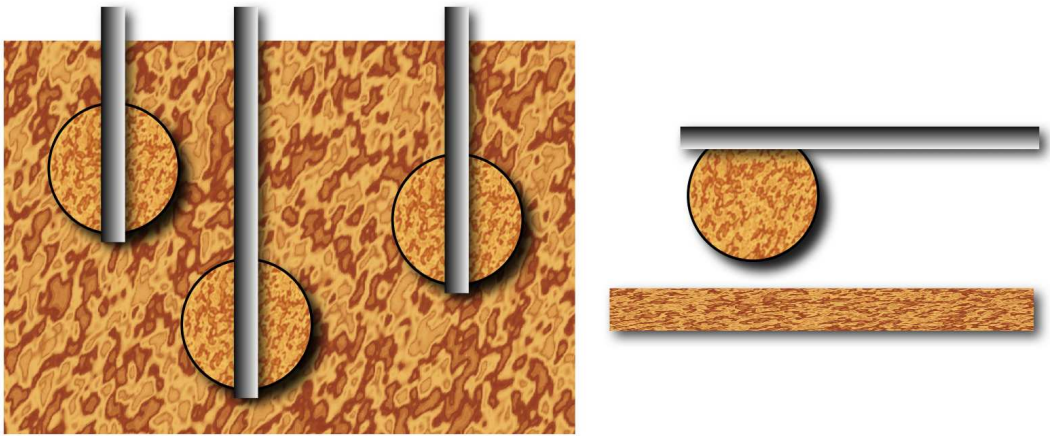}
\caption{ A schematic top view of a spherical AFM tip (right: side view) with disordered charge distribution above a planar substrate with similar charge distribution. Three different realizations of the experiment, i.e. three different lateral positions of the tip above the substrate,  are shown corresponding to three different samples of force data. Each sample would show a different measurement of the normal as well as lateral force with a sample-to-sample variance calculated in the main text.}
\label{schematic}
\end{center}
\end{figure}

To give an example of how these forces could be measured one could take a small randomly charged slab at some distance $l$ from another slab and place its center randomly at some position opposite the larger slab. The same could be done rather more effectively with a plane and a sphere as schematically shown in Fig. \ref{schematic}. Now due to the non-homogeneous quenched charge distribution the smaller slab will experience a random lateral force varying from sample to sample. Such forces could conceivably be readily measurable in an SFA type set up \cite{Israelachvili-r}  used to measure shear forces between solid surfaces sliding past each other across aqueous salt solutions \cite{Raviv} but with interacting surfaces bearing disordered charge distribution. The lateral forces measured in distinct experiments varying in regard to the exact relative lateral positions between the interacting surfaces will average out to zero but we predict that the fluctuations of this lateral force is non-zero and can give information about the magnitude of charge disorder in the system.  The fluctuations we compute here thus correspond to sample-to-sample fluctuations and stem from different sample (experiment) specific relative positions of the interacting surfaces in different experiments. These fluctuations are thus distinct from the temporal fluctuations in the measured force due to thermal fluctuations (an example being thermal fluctuations of the instantaneous thermal Casimir force as discussed in \cite{golest}). 

Most of our computations are for the slab geometry where they can be carried out exactly; however, we show how the lateral force fluctuations can be approximately computed also in the case of the sphere-plane geometry shown on Fig. \ref{schematic}. This configuration is an adaptation of the setup standardly assumed to be within the reach of the {\sl proximity force approximation} (PFA)  \cite{Butt} in the case where the charge disorder on the sphere is assumed to be uncorrelated or very weakly correlated. Using an alternative calculational method where there is no dielectric discontinuity (i.e., all materials used and the intervening space between them have the same dielectric constant), we can compute the lateral force fluctuations exactly  for the sphere-plane system. The form of the PFA developed here agrees with this exact computation in this limit. For the slab geometry we find that the lateral force fluctuations (lateral force variance) behave as  $A/l^2$ where  $A$ is the area of the smaller slab and $l$ is the slab separation. In the sphere-plane set up, within the PFA we find that the lateral force fluctuations behave as $R/l$, where $R$ the radius of  the sphere, and we take the limit where $R\gg l$, with $l$ the closest distance of the sphere to the plane.

For completeness we give the expression for lateral force fluctuations in the case where the intervening medium is  an electrolyte described in the weak-coupling Debye-H\"uckel  approximation \cite{Matej} as well. In this case the force fluctuations are exponentially screened with a screening length given by the Debye length.  

We then turn to the computation of the normal force fluctuations. The method used here is slightly
different as in normal force fluctuations there is a contribution from image charges whose average
is in general non-zero. We reproduce the results of \cite{cd1,cd2} for the average normal force using this method and then go on
to analyze its fluctuations. For the slab geometry with no electrolyte present, we show that the normal force behaves as $A/l^2$, while its variance also scales as $A/l^2$, making both of them comparable.  In the sphere-plane set up,  we find that the fluctuations of the normal force relative to its average value vary  as  $\sqrt{l/R}$  and thus become increasingly more important as 
the separation is increased.

\section{Lateral force fluctuations}

Consider two parallel infinite slabs separated by a distance  $l$. The slabs whose surface is at $z=0$
has a dielectric constant $\epsilon_2$ and the slab whose surface is at $z=l$ has a dielectic constant
$\epsilon_1$. We call these slabs $S_2$ and $S_1$ respectively. We denote by $\epsilon_m$ the dielectric constant of the intervening material. Let each slab have a random surface charge density $\rho_\alpha({\bf x})= \rho_ \alpha({\bf r}, z) $  with zero mean (i.e., the surfaces are net-neutral) 
and correlation  function in the  plane of the slabs  (${\bf r}, {\bf r}'\in S_1, S_2$), i.e.
\begin{equation}
\langle \rho_ \alpha({\bf r}, z) \rho_\beta({\bf r}', z')\rangle=\delta_{\alpha\beta}\,  g_{\alpha s}\,\delta(z-l_\alpha)
\delta(z'-l_\beta)\,C_\alpha({\bf r}-{\bf r}') \qquad  \quad \alpha, \beta=1, 2,  
\end{equation}
and where we define $l_2=0$ and $l_1=l$. In addition we  assume that the charge distribution on slab $S_1$ is restricted to a finite area $A$. In the case where 
the random charge is made up of point charges of signs $\pm e$ of surface density 
$n_{\alpha s}$ then we may write $ g_{\alpha s} = e^2  n_{\alpha s}$, and the correlation function $C({\bf r}-{\bf r'})$ has dimensions of inverse length squared meaning that its two dimensional Fourier transform is dimensionless. Typically the  values of $n_s$ for quite pure samples are smaller than the bulk disorder variance which has a typical range of between $10^{-11}$ to $10^{-6}~ {\rm nm}^{3}$ (corresponding to impurity charge densities of $10^{10}$ to $10^{15}$ $e/\rm{cm}^3$ \cite{cd1}).

The electrostatic energy of the system is given by
\begin{equation}
E = {1\over 2}\int d{\bf x} \,\phi({\bf x})\rho({\bf x})
\end{equation}
where $\rho({\bf x})$ is the total charge density and $\phi({\bf x})$ is the electrostatic potential which 
is given by
\begin{equation}
\phi({\bf x}) = \int d{\bf y} \,G({\bf x},{\bf y})\rho({\bf y})
\end{equation}
while $G({\bf x},{\bf y})$ is the Green's function obeying
\begin{equation}
\epsilon_0\nabla\cdot\epsilon({\bf x}) \nabla G({\bf x},{\bf y}) =-\delta({\bf x}-{\bf y}),
\end{equation}
with $\epsilon({\bf x})$ the local dielectric function. Upon changing the charge distribution the 
corresponding change in the energy of the system is thus given by
\begin{equation}
\delta E = \int d{\bf x} d{\bf y}\,\delta\rho({\bf x}) G({\bf x},{\bf y})\rho({\bf y}).
\end{equation}
If $\rho_1$, the charge distribution on the slab $S_1$, is made up of point charges  we have 
\begin{equation}
\rho_1({\bf x}) = \sum_{i\in S_1}q_i \delta({\bf x}-{\bf x}_i),
\end{equation}
where $q_i$ is the charge at the site ${\bf x}_i$.
Now on moving the smaller slab $S_1$ by a distance ${\bf a}$ laterally, that is to say normally to  the normal between  
slabs, we find that the new charge distribution is simply given by
\begin{equation}
\rho'_1({\bf x}) = \sum_{i\in S_1}q_i \delta({\bf x}-{\bf x}_i-{\bf a}). 
\end{equation}
This means that we can write
\begin{equation}
\delta\rho({\bf x}) = \delta\rho_1({\bf x}) = -{\bf a}\cdot\nabla\rho_1({\bf x}).
\end{equation}
As the plate $S_1$ is moved laterally the self interaction between the charges in both plates is
unchanged, thus the energy change is only given by the interaction of the charges and image
charges in $S_1$ with those in $S_2$. We may thus write
\begin{equation}
\delta E = -{\bf a}\cdot \int 
d{\bf r'}d{\bf r}\,dzdz'\,\nabla_{{\bf r}'}\rho_1({\bf r}',z')G({\bf r}-{\bf r}';z,z')\rho_2({\bf r},z)
\end{equation}
where ${\bf r}'$ and ${\bf r}$ are again the two dimensional coordinates in the planes of $S_1$ and $S_2$ respectively and $z'$ and $z'$ are the respective coordinates normal to the planes. We thus note
that the integration over the coordinate ${\bf r}'$ is over a finite area $A$, while that over ${\bf r}$ is
unrestricted. The lateral force ${\bf F}^{(L)}$ on plate $S_1$ is thus given by 
\begin{equation}
\delta E = -{\bf a}\cdot {\bf F}^{(L)}.
\end{equation}
As the charges in plates $S_1$ and $S_2$ are uncorrelated we find that 
\begin{equation}
\langle \delta E\rangle = -\langle {\bf a}\cdot {\bf F}^{(L)}\rangle = 0,
\end{equation}
that is the average lateral force is zero. 

\begin{figure}[t]
\begin{center}
\includegraphics[angle=0,width=7cm]{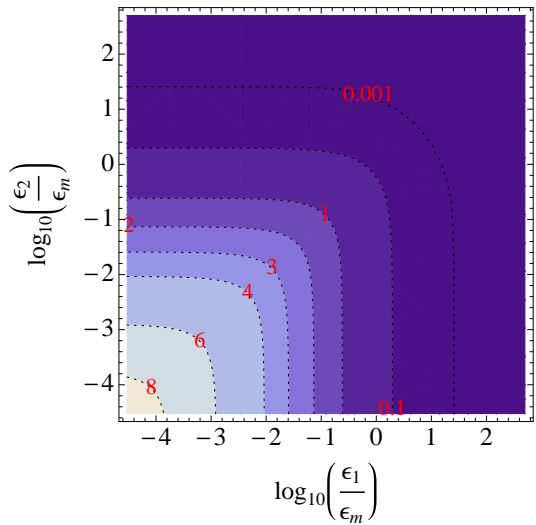}
\caption{ Contour plot of the rescaled lateral force fluctuations,   $f(\epsilon_1/\epsilon_m, \epsilon_2/\epsilon_m)$ (Eqs. (\ref{ncrr}) and (\ref{ncrr2})),  between 
two parallel slabs carrying quenched charge disorder  as a function of $\epsilon_1/\epsilon_m$ and $\epsilon_2/\epsilon_m$ shown here on a $\log_{10}-\log_{10}$ scale.}
\label{fig:f2_lateral}
\end{center}
\end{figure}

The variance of the energy change is given by
\begin{equation}
\langle \delta E^2\rangle 
= a_ia_j\big\langle 
\int d{\bf r'}d{\bf r}dzdz'd{\bf s'}d{\bf s}d\zeta d\zeta'  \,\nabla_{{\bf r}'_i}\rho_1({\bf r}',z')G({\bf r}-{\bf r}';z,z')\rho_2({\bf r},z)
\nabla_{{\bf s}'_j} \rho_1({\bf s}',\zeta')G({\bf s}-{\bf s}';\zeta,\zeta')\rho_2({\bf s},\zeta)\big\rangle, 
\end{equation}
where the summation is over the in-plane Cartesian components $i, j = 1, 2$. As the charge distributions on the two slabs are independent the  only nonzero correlations in the above are given by
\begin{eqnarray}
\langle \rho_2({\bf r},z)\rho_2({\bf s},\zeta)\rangle &=& g_{2s} \delta(z)\delta(\zeta)C_2({\bf r}-{\bf s}) \\
\langle \nabla_{{\bf r}'_i} \rho_1({\bf r}',z')\nabla_{{\bf s}'_j} \rho_1({\bf s}',\zeta')\rangle 
&=& g_{1s}\delta(z'-l)\delta(\zeta'-l)\nabla_{{\bf r}'_i} \nabla_{{\bf s}'_j}C_1({\bf r}'-{\bf s}').
\end{eqnarray}
This then yields 
\begin{equation}
\langle \delta E^2\rangle = a_ia_j\, g_{1s}g_{2s}
\int d{\bf r'}d{\bf r} d{\bf s'}d{\bf s}\, 
G({\bf r}-{\bf r}';0,l)G({\bf s}-{\bf s}';0,l)C_2({\bf r}-{\bf s})\nabla_{{\bf r}'_i}\nabla_{{\bf s}'_j}C_1({\bf r}'-{\bf s}').
\end{equation}
We now write the above in terms of the two dimensional Fourier transforms,
with respect to the in plane coordinates, $\tilde G$ and $\tilde C$ of the functions
$G$ and $C$ and carry out the integrations over the unrestricted coordinates $\bf r$ and $\bf s$  to find
\begin{equation}
\langle \delta E^2\rangle = {a_ia_j g_{1s}g_{2s}\over (2\pi)^4}\int d{\bf k}d{\bf q}\,  d{\bf r}'d{\bf s}'\, 
q_i q_j\, \tilde G(k)^2\tilde C_2(k)\tilde C_1(q)\, \, {\rm e}^{{\rm i}({\bf q}-{\bf k})\cdot({\bf r}'-{\bf s}')},
\end{equation}
where we have used the fact that $\tilde G({\bf k})$ and $\tilde C_i({\bf k})$ are functions of $|{\bf k}|=k$ only.
Now using the fact that the surface charge patch on slab $S_1$ is large (of area $A$) and 
assuming that the correlations between charges are sufficiently short range we may write
\begin{equation}
\langle \delta E^2\rangle = {Aa_ia_jg_{1s}g_{2s}\over (2\pi)^2}\int d{\bf k}\, 
k_i k_j\, \tilde G(k;0,l)^2\tilde C_1(k)\tilde C_2(k) =  {A a^2g_{1s}g_{2s}\over 4\pi }\int dk \, k^3\, \tilde G(k;0,l)^2\tilde C_1(k)\tilde C_2(k),
\end{equation}
where we have used the isotropy of the $k$ integral. From here we deduce that 
 \begin{equation}
\langle F_i^{(L)}F_j^{(L)}\rangle= 
 {A\delta_{ij}g_{1s}g_{2s}\over 4\pi }\int dk \, k^3\, \tilde G(k;0,l)^2\tilde C_1(k)\tilde C_2(k).
 \label{ff1}
 \end{equation}
 The Fourier transform of the Green's function for the parallel slab configuration is easily computed
 by standard methods and is given by
 \begin{equation}
 \tilde G(k;0,l) = {2\epsilon_m \exp(-kl)\over k \epsilon_0(\epsilon_m + \epsilon_1)(\epsilon_m+\epsilon_2)
 (1 -\Delta_1\Delta_2\exp(-2kl))}
 \end{equation}
where
\begin{equation}
\Delta_\alpha = {\epsilon_\alpha - \epsilon_m\over \epsilon_\alpha+\epsilon_m}\quad\qquad \alpha=1, 2, 
\end{equation}
giving the general result in the form
 \begin{equation}
\langle F_i^{(L)}F_j^{(L)}\rangle= 
 {A\delta_{ij}g_{1s}g_{2s} \epsilon_m^2\over \pi\epsilon_0^2 (\epsilon_m + \epsilon_1)^2(\epsilon_m+\epsilon_2)^2}\int k\,dk \, { \exp(-2kl)\tilde C_1(k)\tilde C_2(k)\over (1 -\Delta_1\Delta_2\exp(-2kl))^2}.
 \end{equation}
When the spatial disorder correlations in both slabs are short range such that 
$ C_\alpha({\bf r}-{\bf r}') = \delta({\bf r}-{\bf r}')$, we obtain
\begin{equation}
\langle F_i^{(L)} F_j^{(L)}\rangle =-{A\delta_{ij}g_{1s}g_{2s} \epsilon_m^2\over 4\pi \epsilon_0^2 l^2 (\epsilon_m + \epsilon_1)^2(\epsilon_m+\epsilon_2)^2\Delta_1\Delta_2}\ln(1-\Delta_1\Delta_2), 
\label{ncrr}
\end{equation}
which shows that the lateral force fluctuations decay as $A/l^2$. We may rewrite
this result as 
\begin{equation}
\langle F_i^{(L)} F_j^{(L)}\rangle \equiv-{A\delta_{ij}g_{1s}g_{2s} \over 4\pi  \epsilon_0^2\epsilon_m^2l^2 } \,f\bigg(\frac{\epsilon_1}{\epsilon_m}, \frac{\epsilon_2}{\epsilon_m}\bigg), 
\label{ncrr2}
\end{equation}
where the function $ f(\epsilon_1/\epsilon_m, \epsilon_2/\epsilon_m)$ follows directly from Eq. (\ref{ncrr}). It is plotted in Fig. \ref{fig:f2_lateral} as a function of
$\epsilon_1/\epsilon_m$ and $\epsilon_2/\epsilon_m$.  As can be easily ascertained,  the lateral 
force fluctuations become weaker as $\epsilon_\alpha/\epsilon_m$ tends to infinity (in this case one can see that the function $f(x, y)$ decays, for instance, as $f(x, x)\sim \ln x/x^4$ and $f(x, 0)\sim \ln x/x^2$ when $x\rightarrow \infty$), which corresponds to the case with perfect metallic slabs. On the other hand, when the dielectric constant of the intervening medium is decreased, the force fluctuations become more pronounced and eventually diverge for $\epsilon_\alpha/\epsilon_m\rightarrow 0$ (exhibiting a logarithmic divergence, for instance, as $f(x, x)\sim -\ln x$  when $x\rightarrow 0$). 

The above result means that statistically the lateral force behaves as
\begin{equation}
F_i^{(L)} \sim {\sqrt{A}\over l}.
\end{equation}
Another interesting point here is that the lateral force fluctuations are also present when there are no 
dielectric discontinuities in the system. Here if we set $\epsilon_2=\epsilon_1=\epsilon_m$ we obtain
the result
\begin{equation}
\langle F_i ^{(L)}F_j^{(L)}\rangle ={A\delta_{ij}g_{1s}g_{2s} \over 64\pi  \epsilon_0^2\epsilon_m^2 l^2}.\label{eqpp0}
\end{equation}
This result can be derived in a rather straightforward but illuminating manner that we derive in the Appendix \ref{app}. 

In the case where the intervening medium is composed of an electrolyte with dielectric constant
$\epsilon_m$ and with inverse screening length $m$ in the Debye-H\"uckel approximation we
find that the Green's function obeys
\begin{equation}
\epsilon_0\nabla\cdot\epsilon({\bf x}) \nabla G({\bf x},{\bf y})-\epsilon_0\epsilon({\bf x})\kappa^2({\bf x}) G({\bf x},{\bf y})=-\delta({\bf x}-{\bf y}),
\end{equation}
where as before $\epsilon({\bf x})$ is only a function of $z$ and $\kappa({\bf x})$ is only non-zero (and equal to a constant $\kappa$) within the medium between the two slabs.  From this we obtain
\begin{equation}
 \tilde G(k;0,l) = {2\epsilon_m  K\exp(-Kl)\over \epsilon_0(\epsilon_m K + \epsilon_1k)(\epsilon_m K+\epsilon_2k)
 (1 - \Delta_{1\kappa}\Delta_{2\kappa}\exp(-2Kl))}\label{gfs}
 \end{equation}
 where $K =\sqrt{k^2+\kappa^2}$ and
 \begin{equation}
 \Delta_{\alpha \kappa}= {\epsilon_\alpha k - \epsilon_m K\over \epsilon_\alpha k+\epsilon_m K}, \quad\qquad \alpha=1, 2. 
 \label{deltalab}
 \end{equation} 
In order to obtain the force fluctuations for a system with an intervening electrolyte, at the level
of the Debye-H\"uckel approximation, we simply need to use the expression (\ref{gfs}) in Eq. (\ref{ff1}).

\subsection{PFA for lateral forces}

In many experimental set ups, due to problems of achieving a perfectly parallel alignment, a sphere-plane configuration is used rather than a plane-parallel configuration. In the case where there is
no dielectric discontinuity we can compute the lateral force fluctuations for the sphere-plane geometry.
The derivation given above is easily modified to the case of general geometries if one assumes the validity of the proximity force approximation (PFA) \cite{Butt}. We find in that case that for  the sphere-plane geometry the force correlator is given by 
\begin{equation}
\langle F_i^{(L)}F_j^{(L)}\rangle = {\delta_{ij}g_{1s}g_{2s}\over 32\pi^2\epsilon_0^2\epsilon_m^2}\int dS_1dS_2
{\left({\bf a}\cdot({\bf x}-{\bf y})\right)^2\over \left[ ({\bf x}-{\bf y})^2 +z({\bf x},{\bf y})^2\right]^{3}},
\end{equation}
where ${\bf x}$ are the Cartesian coordinates on surface $S_1$ of object 1 (here the surface of a sphere of radius R) projected onto the in-plane  coordinates of surface 2  and ${\bf y}$ the coordinates on surface  2 or  $S_2$  (here an infinite plate). The variable $z({\bf x},{\bf y})$ is the distance between the  points on the two surfaces perpendicular to the surface $S_2$. In terms of spherical polar coordinates on the surface 
$S_1$ if ${\bf x} = (R\sin \theta\cos \varphi, R\sin \theta\sin \varphi)$ then we have
$z({\bf x},{\bf y})= l + R(1-\cos \theta)$, where $l$ is the distance between the opposing pole of $S_1$ 
and the plane $S_2$  (or the closest distance of the sphere to the plane). The integral can now be written as
\begin{equation}
\langle F_i^{(L)}F_j^{(L)}\rangle = {\delta_{ij}g_{1s}g_{2s}\over 32\pi^2\epsilon_0^2\epsilon_m^2}
\int R^2 \sin \theta \, d\theta d\phi
d{\bf z}\, 
{{\bf z}^2\over \left[ {\bf z}^2 + \big( l + R(1-\cos \theta)\big)^2\right]^{3}},
\end{equation}
where ${\bf z}$ is the relative coordinates of $S_1$ and $S_2$ in the plane of $S_2$ (i.e., it represents ${\bf x}-{\bf y}$ where ${\bf x}$ is in the plane of $S_1$ and ${\bf y}$ in the plane of $S_2$).
Performing the integral over ${\bf z}$ we then find
\begin{equation}
\langle F_i^{(L)}F_j^{(L)}\rangle=  {\delta_{ij}g_{1s}g_{2s}R^2\over 32\epsilon_0^2\epsilon_m^2}\int \sin \theta\,  d\theta 
{1\over  \left( l + R(1-\cos \theta)\right)^2}
\end{equation}
and finally the integral over $\theta$ is easily carried out to give
\begin{equation}
\langle F_i^{(L)}F_j^{(L)}\rangle=  {\delta_{ij}g_{1s}g_{2s}R^2\over 16\epsilon_0^2\epsilon_m^2 l(l+2R)}.\label{exf}
\end{equation}
In the usual experimental set up we are in the limit where $R\gg l$ and we thus find
 \begin{equation}
\langle F_i^{(L)}F_j^{(L)}\rangle\approx  {\delta_{ij}g_{1s}g_{2s}R\over 32\epsilon_0^2\epsilon_m^2 l}.
\end{equation}
In the case where there are dielectric discontinuities we can try to approximate the computation of the force correlator in a manner similar to the proximity force approximation for electrostatic and Casimir 
interaction problems. When the charge distribution are delta-correlated we can assume that the 
force due to the interaction of a unit of area on the sphere at the same separation from the plane
(thus a ring on the sphere) is statistically independent of the others. The ring is specified by the 
polar angle $\theta$ and using Eq. (\ref{ncrr}) we can write that the force on a ring of polar angle between $\theta$ and $\theta+ \delta\theta$ is given by
\begin{equation}
F_i^{(L)}(\theta)= \sqrt{-{g_{1s}g_{2s} \epsilon_m^2\over 4\pi \epsilon_0^2 (\epsilon_m + \epsilon_1)^2(\epsilon_m+\epsilon_2)^2\Delta_1\Delta_2}\ln(1-\Delta_1\Delta_2)}\times {\mu_i(\theta)\sqrt{2\pi R^2\sin \theta\, \delta\theta}\over l + R(1-\cos \theta) },  
\end{equation}
where all the prefactors $\mu(\theta)$ are independent and are of zero mean and variance one. The correlation 
function of the total force is thus given by
\begin{equation}
\langle F_i^{(L)} F_j^{(L)}\rangle = {g_{1s}g_{2s} \delta_{ij}\epsilon_m^2\over 4\pi \epsilon_0^2 (\epsilon_m + \epsilon_1)^2(\epsilon_m+\epsilon_2)^2\Delta_1\Delta_2}\ln(1-\Delta_1\Delta_2) \int {2\pi R^2\sin \theta \, d\theta\over \left[l + R(1-\cos \theta)\right]^2}
\end{equation}
which gives
\begin{equation}
\langle F_i ^{(L)}F_j^{(L)}\rangle = -{\delta_{ij} g_{1s}g_{2s} \epsilon_m^2\over   \epsilon_0^2(\epsilon_m + \epsilon_1)^2(\epsilon_m+\epsilon_2)^2\Delta_1\Delta_2}\ln(1-\Delta_1\Delta_2){R^2\over l(l+2R)},
\end{equation}
and clearly corresponds to the exact result Eq. (\ref{exf}) in the case where there are no dielectric discontinuities.

\section{Normal force fluctuations}

The magnitude of the normal force has been obtained previously \cite{cd1,cd2} and we concentrate our efforts to its fluctuations. The calculations for the normal forces between dielectric slabs with random surface charging are 
slightly different to those above for lateral charges. Here we proceed by writing the electrostatic energy as
\begin{equation}
E = {1\over 2}\int d{\bf x}d{\bf y}\,  \rho({\bf x})G({\bf x},{\bf y}; l)\rho({\bf y})
\end{equation}
where we have made explicit the dependence of the Green's function on the slab separation $l$. The
electrostatic component of the force of the slabs in the normal direction is then given by
\begin{equation}
F^{(N)} = {1\over 2} \int d{\bf x}d{\bf y}\,  \rho({\bf x})H({\bf x},{\bf y}; l)\rho({\bf y}) \qquad 
{\rm where} \qquad  H({\bf x},{\bf y},l) = -{\partial \over \partial l}G({\bf x},{\bf y}; l).
\end{equation}
The average value of the normal force is  non-zero due to the correlation between
the charges in each plate and their image charges \cite{cd1}. In terms of the notations introduced earlier we 
find
\begin{equation}
\langle F^{(N)}\rangle = -{A\over 4\pi}\int k\, dk \, \big[ g_{1s}\tilde H(k;0,0) \tilde C_1(k)
+ g_{2s}\tilde H(k;l,l) \tilde C_2(k)\big].
\end{equation}
The two terms above are the interaction of the charges on surface 1 and 2 with their images. Note that
the contribution of the two surfaces are additive as they are independent. 

In the case where there is electrolyte in the region between the two plates that can be described on the Debye - H\" uckel level one again obtains the relevant expressions for the Green's functions above as
\begin{eqnarray}
\tilde G(k;0,0)&=&{1\over \epsilon_0(K\epsilon_m +k\epsilon_1)}\left({1-\Delta_{1\kappa}\exp(-2Kl)\over 1-\Delta_{1\kappa}\Delta_{2\kappa}\exp(-2Kl)}\right)
\\
\tilde G(k;l,l)&=&{1\over \epsilon_0(K\epsilon_m +k\epsilon_2)}\left({1-\Delta_{2\kappa}\exp(-2Kl)\over 1-\Delta_{1\kappa}\Delta_{2\kappa}\exp(-2Kl)}\right),
\end{eqnarray}
and this then gives the corresponding derivatives $\tilde H$ as
\begin{eqnarray}
\tilde H(k;0,0) &=&  {4K^2\epsilon_m \Delta_{2\kappa}\exp(-2Kl)\over  \epsilon_0
(K\epsilon_m+k\epsilon_1)^2 \left(1-\Delta_{1\kappa}\Delta_{2\kappa}\exp(-2Kl)\right)^2} \\
\tilde H(k;l,l) &=&  {4K^2\epsilon_m \Delta_{1\kappa}\exp(-2Kl)\over  \epsilon_0
(K\epsilon_m+k\epsilon_2)^2 \left(1-\Delta_{1\kappa}\Delta_{2\kappa}\exp(-2Kl)\right)^2} \\
\tilde H(k;0,l) &=&  -{2K^2\epsilon_m \exp(-Kl)\left( 1+ \Delta_{1\kappa}\Delta_{2\kappa}\exp(-2Kl)\right)
\over  \epsilon_0(K\epsilon_m+k\epsilon_1)(K\epsilon_m+k\epsilon_2)\left(1-\Delta_{1\kappa}\Delta_{2\kappa}\exp(-2Kl)
\right)^2}.
\end{eqnarray}
The definition of  $ \Delta_{\alpha \kappa}$ was given in Eq. \ref{deltalab}.  The normal force fluctuations may be computed using Wick's theorem and are given by
\begin{equation}
\langle {F^{(N)}}^2\rangle_c =
{A\over 4\pi}\int k\, dk\,  \big[ g_{1s}^2\tilde H^2(k;0,0)\tilde C^2_1(k) + g_{2s}^2\tilde H^2(k;l,l)\tilde C^2_2(k)
+ 2g_{1s}g_{2s}\tilde H^2(k;0,l)\tilde C_1(k)\tilde C_2(k) \big].
\end{equation}
The first two terms are the force fluctuations due to  the self interactions, i.e. of the charges on surfaces 1 and 2  with their images, and the last term is the fluctuations of the  force  between the charges on surface 1 with those on surface 2 (whose average is always zero).

\begin{figure}[t]
\begin{center}
\includegraphics[angle=0,width=7cm]{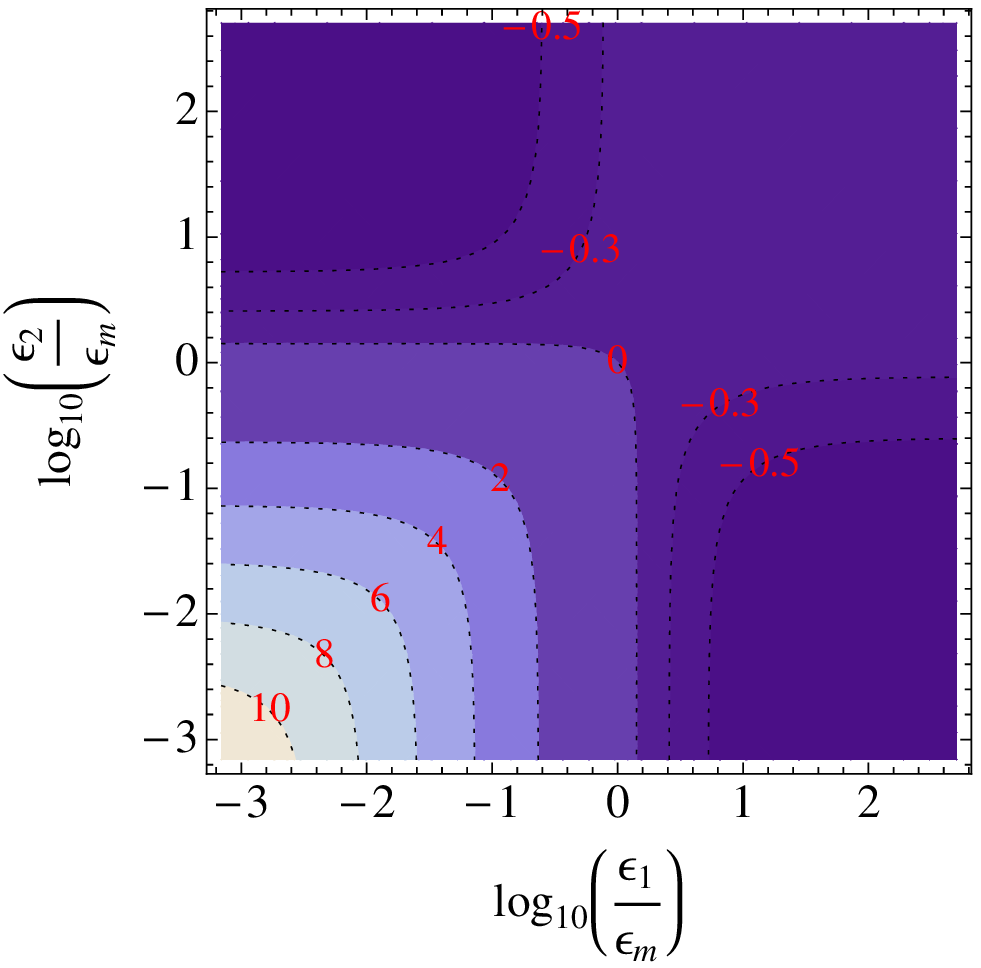}
\caption{ 
Contour plot of the rescaled normal mean force,   ${\cal G}(\epsilon_1/\epsilon_m, \epsilon_2/\epsilon_m, g_{2s}/ g_{1s})$ (Eqs. (\ref{eq:fnormal}) and (\ref{eq:fnormal2})),  between 
two parallel slabs carrying quenched charge disorder with $g_{2s} = g_{1s}$  as a function of $\epsilon_1/\epsilon_m$ and $\epsilon_2/\epsilon_m$ shown here on a $\log_{10}-\log_{10}$ scale.
}
\label{fig:f_normal}
\end{center}
\end{figure}

\begin{figure*}[t]\begin{center}
	\begin{minipage}[b]{0.35\textwidth}\begin{center}
		\includegraphics[width=\textwidth]{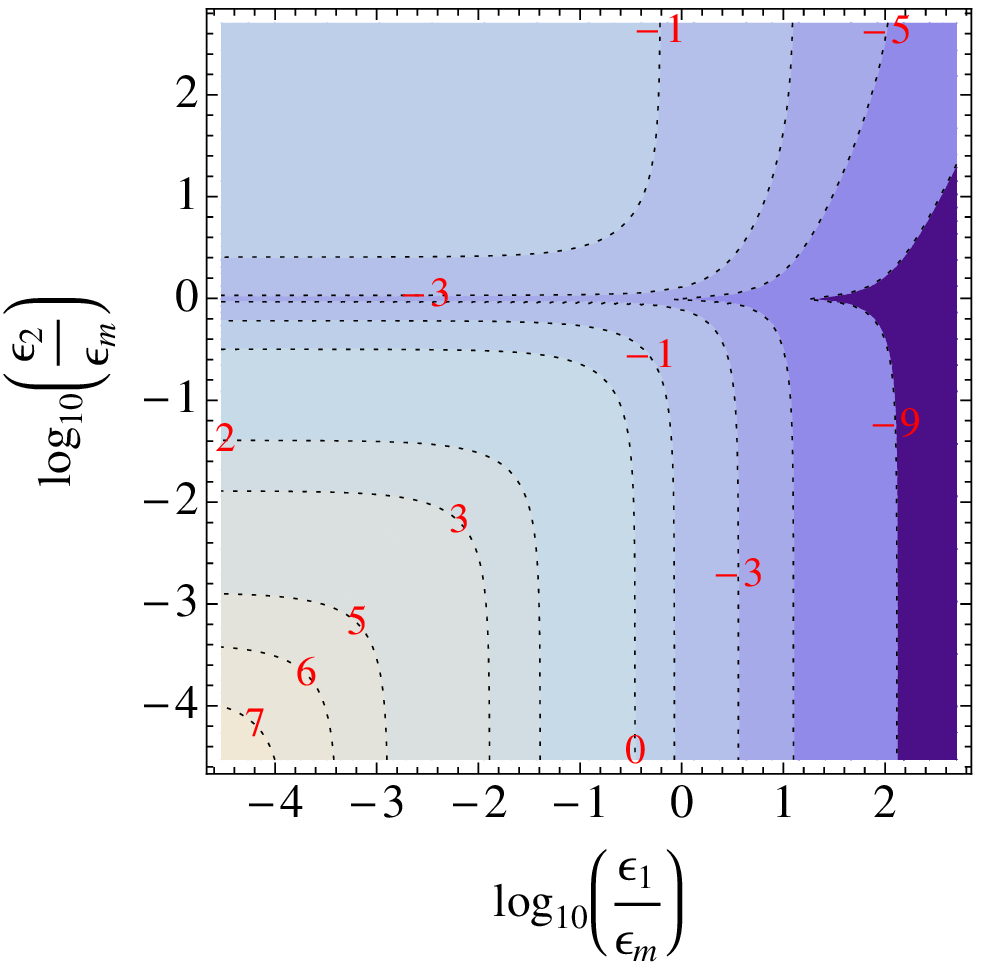} (a)
	\end{center}\end{minipage} \hskip1cm
	\begin{minipage}[b]{0.35\textwidth}\begin{center}
		\includegraphics[width=\textwidth]{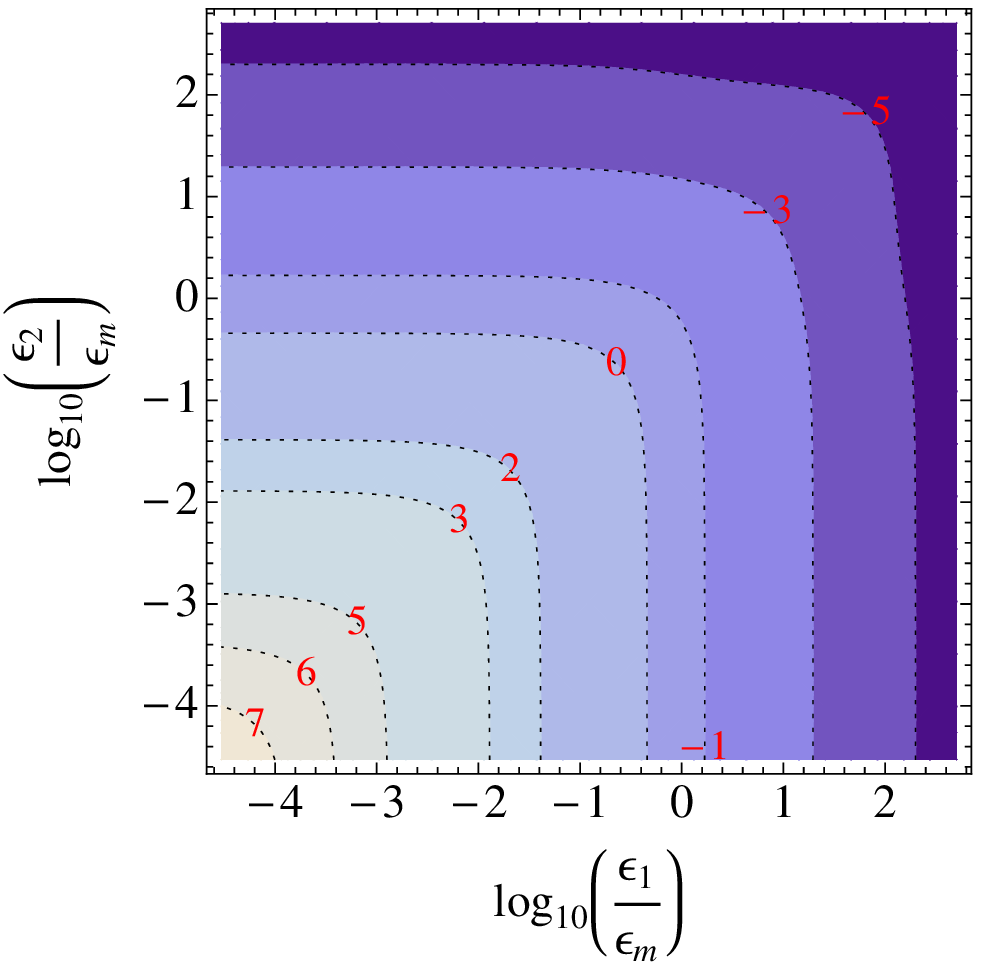} (b)
	\end{center}\end{minipage} \vskip0.3cm
	\begin{minipage}[b]{0.35\textwidth}\begin{center}
		\includegraphics[width=\textwidth]{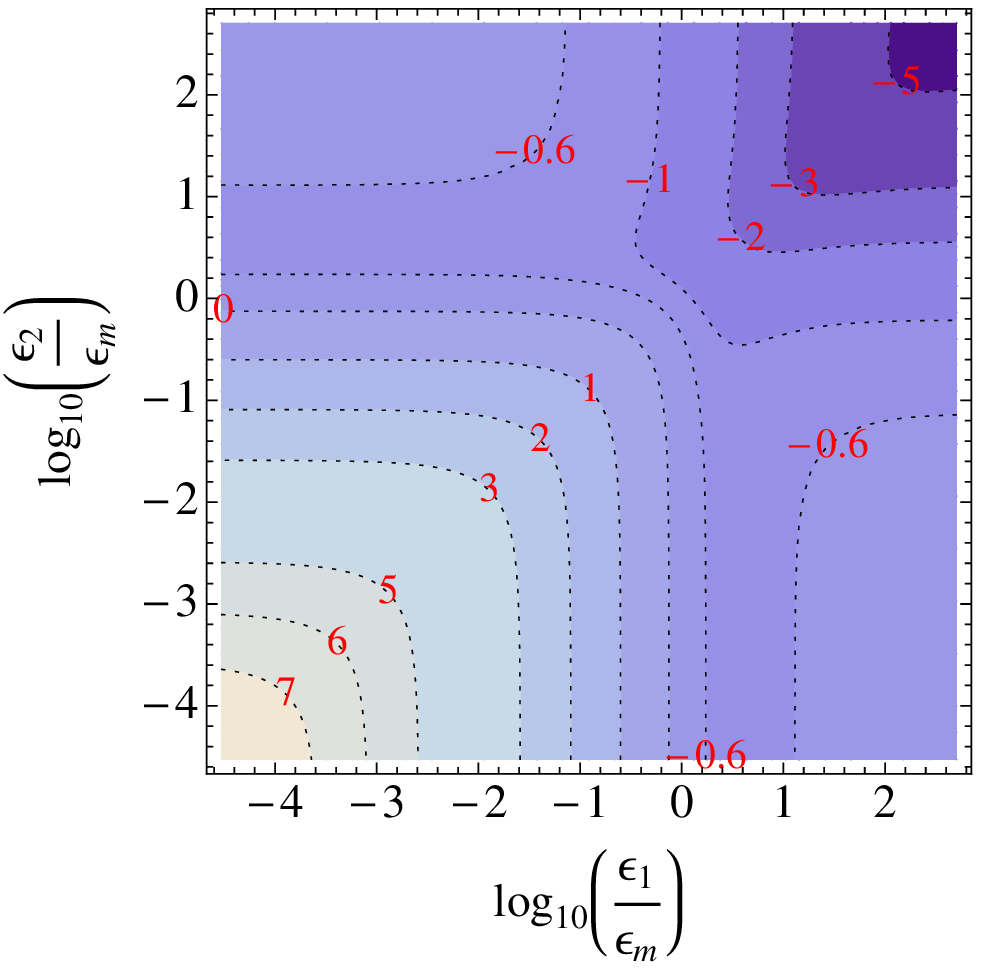} (c)
	\end{center}\end{minipage} \hskip0cm	
\caption{Contour plots of (a) $\log_{10} D_{11}$, the contribution to the force fluctuations due to  the self interactions, i.e. of the charges on surface 1 with their images, and  (b) $\log_{10} D_{21}$,  the contribution to the force fluctuations due to  the interaction between the charges on surface 1 with those on surface 2 as a function of $\epsilon_1/\epsilon_m$ and $\epsilon_2/\epsilon_m$. (c) shows  the rescaled normal force 
fluctuations $\log_{10}{\cal L}\left( \epsilon_1/\epsilon_m, \epsilon_2/\epsilon_m, g_{2s}/ g_{s1}\right)$  for $g_{1s}=g_{2s}$. All plots are shown in  $\log_{10}-\log_{10}$ scale for $\epsilon_1/\epsilon_m$ and $\epsilon_2/\epsilon_m$.
 }
\label{fig:f2_normal}
\end{center}
\end{figure*}

If we take the limiting case where the intervening medium is a simple dielectric devoid of any electrolyte, i.e. $\kappa = 0$,  and where the surface charges are not spatially correlated so that $ C_\alpha({\bf r}-{\bf r}') = \delta({\bf r}-{\bf r}')$, we find that the average value of the normal force is given by
\begin{equation}
\langle F^{(N)}\rangle
= {A\epsilon_m\ln(1-\Delta_1\Delta_2)\over 4\pi  \epsilon_0 l^2}\left( {g_{1s}\over \Delta_1 (\epsilon_m+\epsilon_1)^2}
+{g_{2s}\over \Delta_2 (\epsilon_m+\epsilon_2)^2}\right),
\label{eq:fnormal}
\end{equation}
which recovers our previous results for the average of the normal force due to quenched charge disorder \cite{cd1,cd2}.  This result may be rewritten as 
\begin{equation}
\langle F^{(N)}\rangle
\equiv {Ag_{1s}\over 4\pi \epsilon_0 \epsilon_m l^2}\,{\cal G}\left(\frac{\epsilon_1}{\epsilon_m}, \frac{\epsilon_2}{\epsilon_m},   {g_{2s}\over g_{1s}}\right),
\label{eq:fnormal2}
\end{equation}
where the function ${\cal G}(\epsilon_1/\epsilon_m, \epsilon_2/\epsilon_m, g_{2s}/ g_{1s})$ follows directly from Eq. (\ref{eq:fnormal}) and is shown  in Fig. \ref{fig:f_normal} for the case with $g_{1s}=g_{2s}$. Note that in this case 
the average normal force changes sign and turns from repulsive to attractive when $\epsilon_1/\epsilon_m$ and $\epsilon_2/\epsilon_m$ become larger than
a certain value (shown in the figure by the contour line labeled by 0). For the
symmetric case with $\Delta_1= \Delta_2=\Delta$, one has an attractive force
when $\Delta>0$ (e.g., for two dielectric slabs interacting across vacuum) and a repulsive force when $\Delta<0$. The normal force diverges logarithmically when 
$\epsilon_\alpha/\epsilon_m\rightarrow 0$ as well as when both dielectric
constants $\epsilon_1$ and $\epsilon_2$ tend to infinity (perfect metal limit).
 However, it can take a finite value when only one of the dielectric constants tends to infinity (note, for instance, that ${\cal G}(x, 0)\rightarrow -\ln 2\simeq -0.69$ when
 $x\rightarrow \infty$).

In this case the normal force fluctuations variance 
$\langle {F^{(N)}}^2 \rangle_c = \langle {F^{(N)}}^2 \rangle - \langle {F^{(N)}}\rangle^2$ are given by
\begin{equation}
\langle {F^{(N)}}^2 \rangle_c = {A\over 4\pi\epsilon_0^2 \epsilon_m^2 l^2}\left( g_{1s}^2 D_{11} + g_{2s}^2 D_{22} +
2g_{1s}g_{2s}D_{21}\right)
\end{equation}
where
\begin{eqnarray}
D_{11} &=& {2 \epsilon_m^4 \over 3 (\epsilon_m+\epsilon_1)^4}\left[ {\Delta_2\over \Delta_1(1-\Delta_1\Delta_2)^2} +{\ln(1-\Delta_1\Delta_2)\over \Delta_1^2}\right] \\
D_{22} &=& {2 \epsilon_m^4 \over 3 (\epsilon_m+\epsilon_2)^4}\left[ {\Delta_1\over \Delta_2(1-\Delta_1\Delta_2)^2} +{\ln(1-\Delta_1\Delta_2)\over \Delta_2^2}\right] \\
D_{21}&=& {\epsilon_m^4 \over3  (\epsilon_m+\epsilon_1)^2(\epsilon_m+\epsilon_2)^2}
\left[ -{ 1\over  \Delta_1\Delta_2}\ln(1-\Delta_1\Delta_2) +{2\over  (1- \Delta_1\Delta_2)^2}\right]
\end{eqnarray}
These expressions for $D_{11}$ and $D_{21}$ are 
shown in Figs. \ref{fig:f2_normal}a and b as a function of 
$\epsilon_1/\epsilon_m$ and $\epsilon_2/\epsilon_m$ 
(note that  $D_{22}$
can be obtained from $D_{11}$ by replacing the subindex 1 with 2 and vice versa). In Fig. \ref{fig:f2_normal}c, we show the quantity 
$D_{11}+D_{22}+2D_{21}$ which can be defined as the rescaled normal force 
fluctuations  for the case with $g_{1s}=g_{2s}$  through  
\begin{equation}
\langle {F^{(N)}}^2 \rangle_c \equiv {Ag_{1s}^2\over 4\pi \epsilon_0^2 \epsilon_m^2l^2}\,{\cal L}\left(\frac{\epsilon_1}{\epsilon_m}, \frac{\epsilon_2}{\epsilon_m},   {g_{2s}\over g_{1s}}\right). 
\label{eq:Lfunct}
\end{equation}
The different contributions $D_{\alpha \beta}$ to the normal force fluctuations all 
diverge algebraically  when  $\epsilon_\alpha/\epsilon_m\rightarrow 0$, i.e.,
 ${\cal L}(x, x)\rightarrow x^{-2}$  when $x\rightarrow 0$.

In the case where there are no dielectric discontinuities the forces due to image charges are zero
and the only normal force is due to the interaction between the charges on the two (net-neutral) surfaces (one can easily see that $D_{11}=0$ when $\epsilon_2/\epsilon_m=1$; same is true for
$D_{22}=0$ when $\epsilon_1/\epsilon_m=1$, which explains the 
 non-monotonic behavior of $D_{11}$ and ${\cal L}( \epsilon_1/\epsilon_m, \epsilon_2/\epsilon_m, 1)$ as seen in Fig. \ref{fig:f2_normal}a and c). The mean of the normal force is clearly zero in this case but it has a non-zero variance
\begin{equation}\
\langle {F^{(N)}}^2\rangle_c = {Ag_{1s}g_{2s}\over 32\pi \epsilon_0^2\epsilon_m^2 l^2},
\end{equation}
a result which can be verified using the expression for the Coulomb potential in a system
of constant dielectric constant $\epsilon_m$, with a computation similar to that leading to 
Eq. ({\ref{eqpp}) and then Eq. (\ref{eqpp0}). Interestingly we see, comparing with Eq. (\ref{eqpp0}),
that in the case of a uniform dielectric constant the variance of the force fluctuations in the normal
direction are twice the magnitude as those in the lateral direction.

\subsection{PFA for normal forces}

Within the proximity force approximation for the sphere-plane geometry in complete analogy to the case of lateral force we can derive both the normal force as well as its fluctuations. The former can be obtained in the form
\begin{equation}
\langle F^{(N)}\rangle
= {R^2\epsilon_m\ln(1-\Delta_1\Delta_2)\over   \epsilon_0l(l+2R)}\left( {g_{1s}\over \Delta_1 (\epsilon_m+\epsilon_1)^2}
+{g_{2s}\over \Delta_2 (\epsilon_m+\epsilon_2)^2}\right),
\end{equation}
and the normal force fluctuations as
\begin{equation}
\langle {F^{(N)}}^2 \rangle_c = {R^2\over \epsilon_0^2 \epsilon_m^2 l(l+2R)}\left( g_{1s}^2 D_{11} + g_{2s}^2 D_{22} +
2g_{1s}g_{2s}D_{21}\right).
\end{equation}
From this  formula we see that the relative size of the fluctuations of the normal force to its average
scales as 
\begin{equation}
 {\sqrt{\langle {F^{(N)}}^2\rangle_c}\over \langle F^{(N)}\rangle}\sim \sqrt{{l(l+R)\over R^2}},
\end{equation}
and thus the fluctuations of the normal force relative to its average value become more important as 
the separation is {\em increased} ! The sample-to-sample scatter in the normal force thus increases on increasing the separation between the interacting bodies. This could be interpreted to mean that the interactions themselves restrict their own fluctuations.

\section{Conclusions}

In this work we have proven that the sample-to-sample variance in the lateral as well as normal charge disorder generated forces can be substantial. In the ideal, thermodynamic type, limit where 
the probe area is very large the force is much large than its fluctuations. However in some experimental
set ups the probe size may be quite small and so sample to sample force fluctuations could become
important with respect to average forces. In addition we have shown
that for the sphere plane set up fluctuations become important at large separations where 
the normal force is weak.  In the case of lateral force variance, since the average is, the fluctuations are the only thing remaining. Interestingly enough the fluctuations in the normal and lateral direction are always comparable. For the special case of a uniform dielectric constant we also showed that the variance of the force fluctuations in the normal direction is exactly twice the magnitude of the one in the lateral direction. 

The sample-to-sample variation in the disorder generated force is fundamentally different from the thermal force fluctuations in (pseudo)Casimir interactions as analyzed by Bartolo {\sl et al}. \cite{golest}. In this case one could (in principle at least) use the same experimental setup and just observe the temporal variation of the force at a certain position of the interacting surfaces, measuring the average and the variance within the same experiment (assuming a sufficiently good temporal resolution of the force measuring apparatus). On the other hand, in order to detect sample-to-sample variation one would have to perform many experiments and then look at the variation in the measured force between them. In the first case the variance of the force is intrinsic to the field fluctuations, in the second one it is intrinsic to the material properties of the interacting bodies. 

Additionally, the variance of the fluctuation-induced Casimir force is not universal and is intrinsically related to the  microscopic physics that governs the interaction between the fluctuating (elastic in the case investigated in \cite{golest}) field and the bounding surfaces. 

There are several assumptions in our calculation that need to be spelled out explicitly. We always assume that the position and orientation of the interacting surfaces are fixed in these experiments as well as in the corresponding calculation, just as indicated in the schematic representation of our system on Fig. \ref{schematic}. However, for an unconstrained colloid particle rotational degrees of freedom are not quenched but rather annealed. For example a spherical colloid will rotate so as to minimize its interaction energy in the same way as permanent dipoles orientate with each other. This would introduce additional considerations in the analysis of forces that we do not address in this contribution. In principle there will be random torques  and their sample-to-sample variation can be computed using the methods presented here. These random torques may be accessible to torsion balance based setups \cite{Lambrecht}. 

Additionally, in force measurements slight translations of the sphere with respect the plane and slight rotations of the sphere will lead to different measurements for both normal and lateral forces as well as their fluctuations.

\section{Acknowledgments}

D.S.D. acknowledges support from the Institut Universitaire de France. R.P. acknowledges support from ARRS through the program P1-0055 and the research project J1-0908. A.N. is supported by a Newton International Fellowship from the Royal Society, the Royal Academy of Engineering, and the British Academy. This work was completed at Aspen Center for Physics during the workshop on {\em New Perspectives in Strongly Correlated Electrostatics in Soft Matter}, organized by Gerard C.L. Wong and E. Luijten. We would like to take this opportunity and thank the organizers as well as the staff of the Aspen Center for Physics for their efforts.

\appendix 

\section{Direct calculation of lateral force fluctuations for two slabs with  $\epsilon_2=\epsilon_1=\epsilon_m$}
\label{app}

Let us consider the lateral force fluctuations in the slab system in the absence of  
dielectric discontinuities, i.e. when $\epsilon_2=\epsilon_1=\epsilon_m$. 
The standard three dimensional Coulomb interaction can be written as
\begin{equation}
E = {1\over 2}\int d{\bf x}d{\bf y}\, {\rho({\bf x})\rho({\bf y})\over 4\pi\epsilon_0  \epsilon_m \left[ ({\bf x}-{\bf y})^2 +l^2\right]^{1\over 2}},
\end{equation}
and the change in energy is thus
\begin{equation}
\delta E = \int d{\bf x}d{\bf y}\, {{\bf a}\cdot\nabla \rho_1({\bf x})\rho_2({\bf y})\over 4\pi\epsilon_0 \epsilon_m \left[ ({\bf x}-{\bf y})^2 +l^2\right]^{1\over 2}}.
\end{equation}
The average value of $\delta E$ is clearly zero but one can show that for delta-correlated charge distributions 
\begin{equation}
\langle \delta E^2\rangle = {g_{1s}g_{2s}\over 16\pi^2\epsilon_0^2\epsilon_m^2}\int d{\bf x}d{\bf y}\,
{\left({\bf a}\cdot({\bf x}-{\bf y})\right)^2\over \left[ ({\bf x}-{\bf y})^2 +l^2\right]^{3}}.
\end{equation}
From this one can extract the force correlator as
\begin{equation}
\langle F_i^{(L)}F_j^{(L)}\rangle = {\delta_{ij}g_{1s}g_{2s}\over 32\pi^2\epsilon_0^2\epsilon_m^2}A\int d{\bf z}\,{{\bf z}^2\over ({\bf z}^2 + l^2)^3}, \label{eqpp}
\end{equation}
where the integral over ${\bf z}$ is over the relative position ${\bf x}-{\bf y}$ and the leading order term
is proportional to $A$ (there will be a correction term proportional to the perimeter $\partial A$ of the 
region containing the charge on plate 1).  The integral in Eq. (\ref{eqpp}) is then easily evaluated to recover the result Eq. (\ref{eqpp0}). 


\end{document}